\documentclass[conference]{IEEEtran}
\IEEEoverridecommandlockouts

\usepackage{cite}
\usepackage{amsmath,amssymb,amsfonts}
\usepackage[ruled]{algorithm2e}
\usepackage{graphicx}
\usepackage{textcomp}
\usepackage{xcolor}
\usepackage{hyperref}

\newtheorem{definition}{Definition}

\begin{document}

\title{Locally Differentially Private Embedding Models in Distributed Fraud Prevention Systems
\thanks{Work supported by Innovate UK (project 10048934) through the U.K.-U.S. Privacy-Enhancing Technologies Challenges.}
}

\author{
\IEEEauthorblockN{Iker Perez\IEEEauthorrefmark{1}, Jason Wong\IEEEauthorrefmark{1}, Piotr Skalski\IEEEauthorrefmark{1}, Stuart Burrell\IEEEauthorrefmark{1}, Richard Mortier\IEEEauthorrefmark{2}, Derek McAuley\IEEEauthorrefmark{3} and David Sutton\IEEEauthorrefmark{1}}
\IEEEauthorblockA{\IEEEauthorrefmark{1}Featurespace, United Kingdom}
\IEEEauthorblockA{\IEEEauthorrefmark{2}University of Cambridge, United Kingdom}
\IEEEauthorblockA{\IEEEauthorrefmark{3}University of Nottingham, United Kingdom}
\IEEEauthorblockA{\textit{\{iker.perez, jason.wong, piotr.skalski, stuart.burrell, david.sutton\}@featurespace.co.uk} \\ \textit{rmm1002@cam.ac.uk, derek.mcauley@nottingham.ac.uk}}
}

\maketitle

\begin{abstract}
Global financial crime activity is driving demand for machine learning solutions in fraud prevention. However, prevention systems are commonly serviced to financial institutions in isolation, and few provisions exist for data sharing due to fears of unintentional leaks and adversarial attacks. Collaborative learning advances in finance are rare, and it is hard to find real-world insights derived from privacy-preserving data processing systems. In this paper, we present a collaborative deep learning framework for fraud prevention, designed from a privacy standpoint, and awarded at the recent PETs Prize Challenges. We leverage latent embedded representations of varied-length transaction sequences, along with local differential privacy, in order to construct a data release mechanism which can securely inform externally hosted fraud and anomaly detection models. We assess our contribution on two distributed data sets donated by large payment networks, and demonstrate robustness to popular inference-time attacks, along with utility-privacy trade-offs analogous to published work in alternative application domains.
\end{abstract}


\section{Introduction}

Innovation in Machine Learning (ML) methodologies has made advances to provision financial system actors with tools for fraud prevention and anti-money laundering, which account for losses estimated at 2-5\% of the global Gross Domestic Product \cite{unCrime}. There currently exists an ecosystem of technology partners and risk management vendors that service ML solutions to Financial Institutions (FIs). However, dominant tools are commonly designed for transaction data sets private to each bank or payments processor, which describe money flows only for its account holders. In increasingly cross-border financial flows, FIs are routinely exposed to unfamiliar actors managed by external institutions, and this complicates fraud assessments according to independent evaluations \cite{FATF}. Consequently, governmental bodies are working to produce incentives for financial data sharing, and financing research contributions and innovation competitions that explore collaborative ML approaches and \textit{Privacy Enhancing Technologies} (PETs) \cite{PETsChallenge}.

Simultaneously, fraud prevention systems have transitioned from rules-based algorithms and tree-based classification models towards hybrid processes that leverage Deep Learning (DL) architectures for sequential data streams. These are increasingly researched and deployed in controlled production settings \cite{roy2018deep, branco2020interleaved, wong2023training} and often rely on Recurrent Neural Networks \cite{hochreiter1997long, yu2019review, 9912385} or Attention mechanisms \cite{vaswani2017attention}. Integrating DL within fraud prevention reduces the need for extensive feature engineering \cite{lecun2015deep} and facilitates the retrieval of \textit{embedded} representations for financial transactions, which capture fraud typologies in a manner reusable across systems. However, DL model deployments may memorise and unintentionally leak personal or commercially sensitive information when exposed to partnering institutions and external parties \cite{shokri2015privacy, nasr2018comprehensive, carlini2019secret, zhu2019deep, yin2021comprehensive}. 

Privacy leaks may compromise both training data and inference-time inputs. Commonly, attackers target public interfaces that expose inference-time functionalities for distributed models, however, these may also be susceptible to training-time attacks by malicious parties in a consortium \cite{mothukuri2021survey, yin2021comprehensive}. Attack typologies are diverse, and popular choices include membership, attribute inference and model inversions. In all cases, a model becomes susceptible to attacks due to external over-exposure to its logic, architecture, gradients, predictive confidence or latent representations \cite{lyu2020threats}. Thus, there exists a growing body of work exploring model design, training and deployment from privacy standpoints across a variety of domains including text, vision or speech, and considering multiple threat models and adversarial exploitation strategies \cite{coavoux2018privacy, mahendran2015understanding, ulyanov2018deep, song2020information}. Also, we observe advances in privacy-preserving DL through techniques including differential privacy \cite{dwork2008differential, abadi2016deep}, homomorphic encryption \cite{aono2017privacy} or secure multi-party computation \cite{knott2021crypten}. Unfortunately, it is hard to find evidence-based studies derived from data processing systems in finance, due to confidentiality requirements and a lack of open-sourced data sets with distributed sources of payments.

In this paper, we introduce a novel privacy-preserving collaborative learning framework for fraud and anomaly detection, awarded at the recent PETs Prize Challenges \cite{PETsChallenge} hosted by the U.K. \textit{Centre for Data Ethics and Innovation}, \textit{Innovate UK}, the U.S. \textit{National Institute of Standards and Technology}, and the \textit{National Science Foundation} (NSF), in collaboration with the \textit{SWIFT} global payments network \cite{SwiftSite}. The proposed setting aligns with vertical Federated Learning (FL) scenarios \cite{liu2020asymmetrical, romanini2021pyvertical}, where participating institutions do not share easily exploitable model architectures \cite{melis2019exploiting}. We present a data publication mechanism for transaction histories of private accounts, using embedded representations endowed with Local Differential Privacy (LDP), and we discuss distributed training scenarios through pipeline parallelism and back-propagation. We present evidence of good utility-privacy trade-offs with three classes of attacks on two distinct applications, using large synthetic and real-world distributed transaction data provided by SWIFT and Featurespace \cite{Featurespace} for research purposes.

\noindent\textbf{Statement of Contributions}. All in all, we present fraud prevention and privacy-preserving concepts in a winning solution to the PETs challenge \cite{SwiftSite}, along with processes for training a distributed system. In doing so, we:
\begin{itemize}
    \item Formalise a privacy-preserving release mechanism for sequential transaction data through latent embeddings. 
    \item Explore vulnerabilities to common inference-time attacks. 
    \item Publish a privacy analysis with LDP in financial data.
\end{itemize}
We discover that fraud models processing large quantities of financial data yield utility-privacy profiles analogous to outputs in alternative application domains \cite{mahendran2015understanding, coavoux2018privacy, ulyanov2018deep, song2020information}. With suitable precautions, embedded representations published by data owners do not encode comprehensive characteristics or behaviours of account holders, thus, inversion, inference or membership attacks show limited success rates. The rest of the paper is organised as follows. In Section \ref{sec:FraudInto} we review fraud detection and collaborative fraud prevention methods, while \ref{sec:EmbeedingsAndPrivacy} summarises results for embedding models, differential privacy and additive mechanisms. In \ref{sec:main} we present a privacy-preserving collaborative framework for fraud prevention, along with algorithms for distributed training. Finally, sections \ref{sec:Experiments} and \ref{sec:Conclusion} contain synthetic and real-world experimentation results, along with a discussion.

\section{Fraud Detection in Financial Transactions and Payments} \label{sec:FraudInto}

Fraud detection systems are deployed by banks and payments processors alike, in order to monitor sequences of transaction data with high throughput and low-latency. The institutions are generally liable for losses accrued by account holders in transactions executed without their authorisation. Thus, a model aims to estimate the likelihood of fraud associated with a transaction, in order to trigger additional authentication procedures. In its simplest setting, this is formulated as a supervised learning task for binary classification \cite{bolton2002statistical}. Here, a transaction $\boldsymbol{x}_t\in\mathcal{X}$ recorded at time $t>0$ is endowed with an \textit{anomaly} or \textit{fraud} label $y\in\{0,1\}$, and conditional responses are considered Bernoulli distributed, s.t. $$y|\boldsymbol{x}_t\sim\text{Ber}(\mathbb{P}(y=1|\boldsymbol{X}_{\leq t}))$$ where $\boldsymbol{X}_{\leq t}$ denotes the set of all transactions preceding and including $\boldsymbol{x}_t$. Labels are procured using payment dispute information, such as \textit{chargebacks}, and are scarce relative to overall transaction volumes. The support $\mathcal{X}$ accommodates numerical and categorical variables, including monetary values and transaction times, as well as reference codes for merchant types, currencies or accounts; we refer the reader to \cite{iso, iso2} for schema types in financial messaging. 
Finally, note that transactions originate at an \textit{ordering} account and credit a \textit{beneficiary}, both drawn from a pool of uniquely identifiable IDs for merchants and payment cards. These identifiers are always recorded in financial messages.

\vspace{4pt}
\noindent\textbf{Remark.} \textit{For simplicity, we assume that transaction data in $\mathcal{X}$ is numerically formatted. This may be achieved through common pre-processing and tokenisation steps} \cite{hancock2020survey}.
\vspace{6pt}

A regression function $f:\cup_{n=0}^\infty\mathcal{X}^n \rightarrow [0, 1]$ is thus designed to estimate $\mathbb{P}(y=1|\boldsymbol{X}_{\leq t})$, and trained on tuples $(\boldsymbol{X}_{\leq t}, y)$ of transaction sequences with fraud labels. Its output is referred to as a \textit{score} \cite{bolton2002statistical}, and used by downstream decision engines for rejecting transactions based on risk appetites. Above, $\mathcal{X}^n$ is the $n$-fold product space of $\mathcal{X}$ and $\cup_{n=0}^\infty\mathcal{X}^n$ its union; this signifies that $f$ takes as input sequences of arbitrary length.

\subsection{Decomposing state-of-the-art classifiers} \label{sec:modelDef}

\begin{figure*}[h!]
    \centering
    \includegraphics[width=0.94\textwidth]{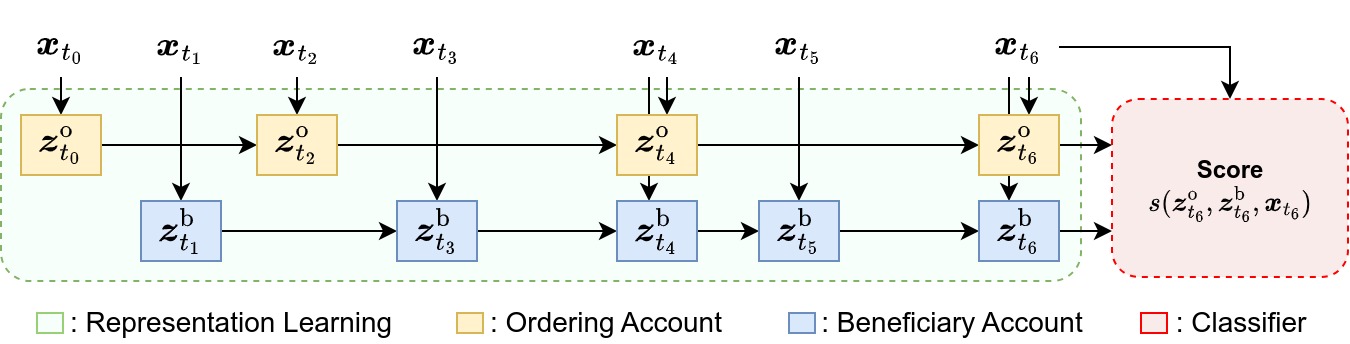}
    \caption{Schematic of a modern fraud classifier. To score a transaction $\textbf{x}_{t_6}$ at time $t_6$, historic data for ordering and beneficiary accounts is ingested, formatted and independently aggregated over time. The resulting profiles $\textbf{z}^\text{o}_{t_6}$ and $\textbf{z}^\text{b}_{t_6}$ are processed by a binary classifier, producing a score.}
    \label{fig:decomposing}
\end{figure*}

A fraud classifier must address the sequential structure of the problem. Historic events are key to understanding typical financial behaviours for transacting accounts, and to identify any anomalies. The building blocks of a modern classifier as pictured in Figure \ref{fig:decomposing} are:
\begin{itemize}
    \item A \textbf{pre-processing} layer to format and filter historic events for ordering and beneficiary accounts in a transaction.
    \item A \textbf{representation learning} module $\psi:\cup_{n=0}^\infty\mathcal{X}^n \rightarrow \mathbb{R}^m$ to discover features from historic events for ordering and beneficiary accounts.
    \item A \textbf{regressor} $s:\mathbb{R}^m\times \mathbb{R}^m\times \mathcal{X}\rightarrow [0, 1]$ to map account features and the current transaction into scores.
\end{itemize}
Thus, a historic sequence of transactions
\begin{equation}
\boldsymbol{X}_{\leq t}=\{\boldsymbol{x}_{t_i} \ : \ i=0,1,\dots \ s.t. \ t_i\leq \ t\} \label{eq:Sequence}
\end{equation}
at times $\{t_i\}_{i=0, 1, \dots}$ and leading to $\boldsymbol{x}_t$ is filtered into sub-sequences $\boldsymbol{X}^{\text{O}}_{\leq t}$ and $\boldsymbol{X}^{\text{B}}_{\leq t}$ for the \textit{ordering} and \textit{beneficiary} accounts. A transaction score is then procured as $$f(\boldsymbol{X}_{\leq t}) = s(\psi(\boldsymbol{X}^{\text{O}}_{\leq t}), \psi(\boldsymbol{X}^{\text{B}}_{\leq t}), \boldsymbol{x}_t),$$
commonly using a \textit{multi-layer perceptron} \cite{lecun2015deep} as  regressor. The latent feature vector $\boldsymbol{z}^{\text{o}}_{t}=\psi(\boldsymbol{X}^{\text{O}}_{\leq t})$ (similarly defined for the beneficiary) is referred to as an \textit{embedded} transaction profile at time $t>0$, and is reusable across fraud detection systems \cite{weiss2016survey}.

Representation learning processes in $\psi$ may be replaced with manual feature engineering. However, these typically rely on differentiable recurrent architectures that yield embedded profiles \cite{9912385}, for example:
\begin{align}
    \boldsymbol{z}_{t_0} &= \sigma(W\boldsymbol{x}_{t_0} + b), \nonumber\\
    \boldsymbol{z}_{t_i} &= \sigma(W\boldsymbol{x}_{t_i} + U\boldsymbol{z}_{t_{i-1}}+b) \ \text{for} \ i=1,2, \dots \nonumber\\
    \psi(\boldsymbol{X}_{\leq t}) &= \boldsymbol{z}_{t} \label{eq:embedding}
\end{align}
where $W$, $U$ and $b$ are learnable weight tensors and $\sigma$ is an arbitrary non-linear activation function. However, \textit{long short-term memory} and \textit{gated recurrent unit} formulations \cite{yu2019review} are the norm, along with self-attention mechanisms \cite{vaswani2017attention}. We note that it is challenging to accommodate in-homogeneous inter-arrival times for transactions in formulae \eqref{eq:embedding} and we refer the reader to \cite{li2022approximation} for a discussion on this topic.

\subsection{Privacy-preserving collaborative fraud prevention}

Reliable feature profiles for card and merchant accounts in \eqref{eq:embedding} can only be generated from complete transacting histories, available only to their managing FIs. When a bank or payments processor deploys a fraud detection system, this must approximate profiles for externally managed accounts, using partial data observable through customer interactions across institutions. Yet, advances in Federated Learning (FL) \cite{bonawitz2019towards, li2020federated, zhang2021survey} and privacy technologies \cite{dwork2008differential, naehrig2011can, dwork2014algorithmic, heurix2015taxonomy} have unlocked collaborative learning solutions \cite{dankar2013practicing, wang2016differential, PETsChallenge}. FL frameworks aim to train collaborative models on heterogeneous data sets distributed across owners, exchanging parameters or gradients through an orchestrator \cite{zhang2021survey}. This is problematic; models are known to leak information about data sources when subjected to adversarial attacks \cite{zhu2019deep, melis2019exploiting}. Thus, complementary PETs are usually utilised by data owners when publicizing model updates \cite{abadi2016deep}, or integrated in federated learning designs that restrict susceptibility to malicious attacks \cite{aono2017privacy, knott2021crypten, xu2021privacy}.

In finance and payments, design choices in fraud models are influenced by regulations pertaining personal data, such as the EU \textit{General Data Protection Regulation}. There further exist concerns around the operational costs, complexity and risk of enabling privacy-preserving federated learning solutions, due to fears of hidden vulnerabilities and scalability problems \cite{darbha2020privacy}. There exist few methodological advances or analyses on real deployments \cite{almuhammadi2004better, kanamori2022privacy, canillas2018exploratory, lin2020protecting}. Thus, we present a novel strategy for privacy-preserving data publication, along with a collaborative framework for training fraud models as described in subsection \ref{sec:modelDef}. We describe a means for FIs to securely publish complete embedded profiles for accounts in \eqref{eq:embedding}, and to leverage them in external downstream ML fraud classifiers. 

\section{Embedding Models and Privacy} \label{sec:EmbeedingsAndPrivacy}

Embeddings are de-facto building blocks in advanced DL applications, due to their flexibility and reusability \cite{gui2016large, reimers2019sentence, song2020information}. They are defined as numerical representations of categorical arrays or high-dimensional feature vectors, such as word tokens, images or audio files. These retain important information by organising sparse inputs into an \textit{embedding space}, s.t. similar inputs lie in proximity, and can be repurposed in downstream predictive tasks. 

\begin{figure}[h!]
    \centering
    \includegraphics[width=0.46\textwidth]{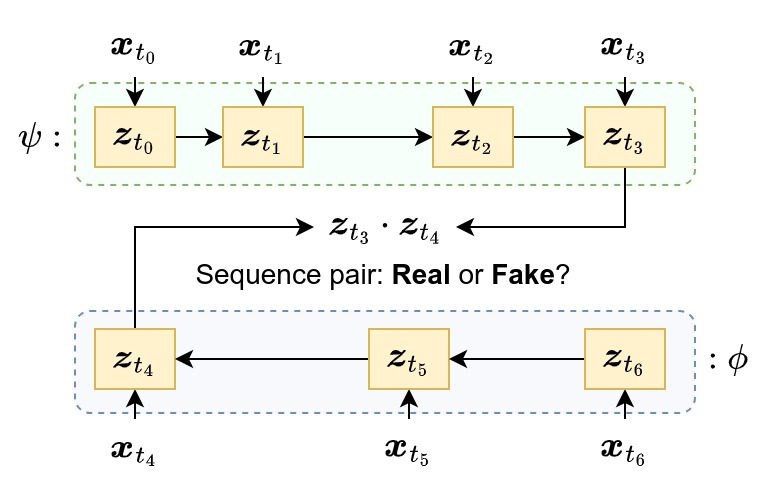}
    \caption{Dual encoding procedure for transaction sequences. Forward and backward embeddings are combined into a contrastive loss function.}
    \label{fig:contrastive}
\end{figure}

\noindent\textbf{Transaction embeddings}. In financial applications, embeddings may compress large quantities of variable-length transaction data into behaviour profiles. A recurrent representation learning function as introduced in \eqref{eq:embedding} acts as an embedding function, s.t. sequences $\boldsymbol{X}_{\leq t}$ in \eqref{eq:Sequence} are mapped into $m$-dimensional vectors $\boldsymbol{z}_t=\psi(\boldsymbol{X}_{\leq t})\in\mathbb{R}^m$ with information usable by a fraud classifier. This may be trained alongside a scoring function in an end-to-end pipeline for fraud detection. However, it is desirable to capture comprehensive representations of inputs. Thus, embeddings are often trained using unsupervised mechanisms for \textit{next-event} prediction, \textit{meet-in-the-middle} and \textit{dual encoding} \cite{logeswaran2018an, reimers2019sentence, nguyen2023meet}. For instance, we use historic sequences $\boldsymbol{X}_{\leq t}$ for accounts, leading to both real $\boldsymbol{X}_{>t}$ and \textit{negatively sampled} $\{\boldsymbol{X}^-_{k, >t}\}_{k=1,\dots,K}$ future transactions, and we minimise a contrastive loss \cite{babaev2022coles}
\begin{equation}
-\log\frac{\exp \big(\boldsymbol{z}_{t} \cdot \boldsymbol{z}_{>t}/\tau\big)}{\sum_{k=1}^K \exp\big(\boldsymbol{z}_{t} \cdot \boldsymbol{z}^-_{k,>t}/\tau\big)} \label{eq:Contrastive}
\end{equation}
with arbitrary temperature $\tau>0$, where $\boldsymbol{z}_{>t}=\phi(\boldsymbol{X}_{>t})$ is a \textit{backwards} embedding, procured with a recurrent network $\phi\approx\psi$ which processes future transactions in reversed order, as shown in Figure \ref{fig:contrastive}.

FIs may thus share profiles for accounts under their management, and help models hosted by external partners. However, embeddings have shown vulnerabilities against a range of privacy attacks, including inversion, membership and attribute inference \cite{song2020information}. This demands further privacy guarantees to conceal individuals.

\subsection{$\epsilon$-Local Differential Privacy}

Differential privacy (DP) and local differential privacy (LDP) refer to randomised-response mechanisms for data querying, release and publication, which disclose useful patterns while withholding detailed descriptions \cite{dwork2014algorithmic}.

\begin{definition}[$\epsilon$-differential privacy]
A randomised data publication mechanism $\mathcal{M}$ fulfills $\epsilon$-DP, if for any two databases $D$ and $D'$ differing on a single data element
$$\mathbb{P}(\mathcal{M}(D)\in S) \leq e^\epsilon \cdot \mathbb{P}(\mathcal{M}(D')\in S)$$
for all subsets $S\in\textnormal{Im}(\mathcal{M})$ and $\epsilon>0$, with probabilities defined over randomness in $\mathcal{M}$.
\end{definition}

Intuitively, $\mathcal{M}$ is an algorithm with user-defined aggregation logic. It takes arbitrary input data and outputs randomised values. The definition above imposes a probabilistic constraint on the dissimilarity of outputs for different inputs $D$ and $D'$, determined by a privacy \textit{budget} denoted $\epsilon$. When $\epsilon\rightarrow 0$ all outputs are distributionally indistinguishable, and we may not infer whether a particular element in the data was used during aggregations. This process protects individual privacy in queries over large populations. However, if private transaction data pertaining a single account is aggregated and published, then a restrictive local definition of DP is required \cite{wainwright2012privacy}.
\begin{definition}[$\epsilon$-local differential privacy]
A randomised user data publication mechanism $\mathcal{M}$ fulfills $\epsilon$-LDP, if for any two pairs of possible private data $\boldsymbol{z}$ and $\boldsymbol{z}'$, we have
$$\mathbb{P}(\mathcal{M}(\boldsymbol{z})\in S) \leq e^\epsilon \cdot \mathbb{P}(\mathcal{M}(\boldsymbol{z}')\in S)$$
for all subsets $S\in\textnormal{Im}(\mathcal{M})$ and $\epsilon>0$, with probabilities defined over randomness in $\mathcal{M}$.
\end{definition}

Here, we constrain the dissimilarity of outputs produced by data in any two individual accounts. For further properties associated with DP mechanisms, such as composability or robustness to post-processing, we refer the reader to \cite{dwork2014algorithmic, dwork2008differential}. Overall, these tools offer robust privacy guarantees, and are commonly used to protect healthcare, financial, geo-location or survey data, across a range of systems used for recommendation, regression and processing or text, images and audio \cite{phan2017adaptive, wang2016differential, abadi2016deep, dankar2013practicing, lin2020protecting}.

\subsection{Additive Noise Mechanisms}

Additive methods are simple randomised response procedures amenable to numerical outputs. Here, a DP mechanism is designed by altering user data with noise drawn from \textit{Laplace} or \textit{Gaussian} distributions \cite{phan2017adaptive}. In the simplest setting \cite{dwork2006calibrating}, an $\epsilon$-LDP mechanism $\mathcal{M}:\cup_{n=0}^\infty\mathcal{X}^n \rightarrow \mathbb{R}^m$ for publishing account transaction profiles, is constructed as a function composition $\mathcal{M} = \mathcal{M}^{\epsilon}_{\text{Laplace}}\circ \psi$, where $\psi$ is a representation learning model as defined in \eqref{eq:embedding}, and
\begin{equation}
\mathcal{M}^{\epsilon}_{\text{Laplace}}(\boldsymbol{z})=\boldsymbol{z}+(w_1, \dots, w_m) \label{eq:mechanism}
\end{equation}
with $w_i\sim\text{Lap}(\Delta\psi/\epsilon)$ for $i=1,\dots,m$. Thus we add noise to embeddings $\boldsymbol{z}=\psi(\boldsymbol{X}_{\leq t})$ using independent $0$-centred Laplace distributions with scale $\Delta\psi/\epsilon$, where the \textit{sensitivity} of the embedding function is defined as
$$\Delta\psi = \max_{\boldsymbol{X}_{\leq t} , \boldsymbol{X}'_{\leq t}}\lVert \psi(\boldsymbol{X}_{\leq t}) - \psi(\boldsymbol{X}'_{\leq t}) \rVert_1$$
i.e. the maximum $\ell_1$-norm over all transaction sequence pairs $\boldsymbol{X}_{\leq t} , \boldsymbol{X}'_{\leq t}\in\cup_{n=0}^\infty\mathcal{X}^n$ observed across all accounts. The sensitivity quantifies the upper bound on how much embedded profiles may vary across different transaction sequences. In practice, this is unfeasible to estimate, as it requires evaluation across infinitely varied sequences. Instead, the architecture of $\psi$ is intentionally designed s.t. its output embedding space is \textit{bounded} in $\ell_1$-norm, using \textit{clipping} to guarantee $\Delta\psi\leq 1$. We refer the reader to \cite{dwork2014algorithmic} for relevant privacy proofs, along with further details for constructions of advanced additive noise mechanisms, such as \textit{Gaussian} mechanisms.

\section{Private Embedding Models for Distributed Anomaly and Fraud Detection} \label{sec:main}

We describe distributed training strategies with model formulations as introduced in subsection \ref{sec:modelDef}. Here, a payments processor or bank $\beta=1,\dots,B$ hosts a regressor function $s_\beta:\mathbb{R}^m\times \mathbb{R}^m\times \mathcal{X}\rightarrow [0, 1]$ to score transactions $\boldsymbol{x_t}\in\mathcal{X}$ under its management. This ingests profiles $\boldsymbol{z}^{\text{o}}_{t}, \boldsymbol{z}^{\text{b}}_{t} \in \mathbb{R}^m$ for ordering and beneficiary accounts, aggregated up to time $t>0$, which may be of two types:
\begin{itemize}
    \item Embeddings $\boldsymbol{z}_{t}=\psi_\beta(\boldsymbol{X}_{\leq t})$ for accounts managed by a bank $\beta$, using a private representation learning function $\psi_\beta:\cup_{n=0}^\infty\mathcal{X}^n \rightarrow \mathbb{R}^m$.
    \item Locally differentially private embeddings $\boldsymbol{z}_{t}=\mathcal{M}(\boldsymbol{X}_{\leq t})$ for external accounts, published by partner institutions using additive mechanisms as described in \eqref{eq:mechanism}.
\end{itemize}
Thus, profiles are aggregated on complete histories only available to their managing institutions. We continue with two popular scenarios in payments. First, a peer-to-peer setting where FIs forward money to each other through real-time payment services, detached from fraud validation processes. Second, an orchestrated system where a messaging provider, such as SWIFT, acts as a centralised transacting channel for external institutions.

\subsection{Transfer Learning with Pre-Trained Embeddings}

Institutions that use machine learning models for fraud prevention store up-to-date profiles for all of their accounts, either manually engineered or embedded in unsupervised settings, as described in \ref{eq:Contrastive}. In a peer-to-peer scenario, we leverage transfer learning \cite{weiss2016survey} through profiles that already exist. Here, a bank or FI trains a scoring function by taking as input internal and externally published profiles. A naive mini-batch step with \textit{stochastic gradient descent} is summarised in Algorithm \ref{alg:peer} for bank $\beta=1$. Without loss of generality, this assumes that ordering accounts are always internal, while beneficiary profiles are requested from partners, which use independent representation learning and privacy mechanisms $\{\psi_\beta, \mathcal{M}_\beta\}_{\beta=2,\dots, B}$. Thus, external profiles are in-homogeneous and require pre-processing using learnable functions $g_{1,\beta}:\mathbb{R}^{m_\beta}\rightarrow \mathbb{R}^m$ across peers $\beta=2,\dots,B$. The algorithm scales linearly with a complexity of $O(B N \omega)$ in a mini-batch of size $N$, where $w$ denotes the maximum number of weights across all scoring function, pre-processors and data-release mechanisms.
\begin{algorithm}[h!]
\caption{SGD step. Bank $\beta=1$. Peer-to-peer.}\label{alg:peer}
\KwIn{Mini-batch $\{(\boldsymbol{x}_{i, t_i}, \boldsymbol{z}^{\text{o}}_{i, t_i}, y_i)\}_{i=1,\dots, N}$. \newline Loss function $\mathcal{L}$ and learning rate $\lambda>0$. \newline Slack term $\gamma\rightarrow0^+$.
}
\KwOut{Updated $s_1$ and $g_{1,\beta}$, $\beta=2,\dots, B$.}
\For{$i=1, \dots, N$}{
    $\rho_i \gets getBeneficiary(\boldsymbol{x}_{i, t_i})$ \\
    $\hookrightarrow$ \hfill \textbf{Beneficiary Bank } $\rho_i$ \\
    \hfill Publish $\boldsymbol{z}^{\text{b}}_{i, t_i} = \mathcal{M}_{\rho_i}(\boldsymbol{X}^{\text{B}}_{\leq t_i})$ \\
    Pre-process $\boldsymbol{r}^{\text{b}}_{i, t_i} = g_{1,\rho_i}(\boldsymbol{z}^{\text{b}}_{i, t_i})$ \\
    Predict $\hat{y}_i = s_1(\boldsymbol{z}^{\text{o}}_{i, t_i}, \boldsymbol{r}^{\text{b}}_{i, t_i}, \boldsymbol{x}_{i, t_i})$ 
}
Update weights $\omega_s$ of $s_1$:
\begin{equation*}
    \omega_s \gets \omega_s - \frac{\lambda}{N}\sum_{i=1}^N \nabla_{\omega_s}\mathcal{L}\big(y_i, \hat{y}_i \big)
\end{equation*}
\For(update weights $\omega_{g}$ of $g_{1,\beta}$:){$\beta=2,\dots, B$}{
\begin{equation*}
    \omega_{g} \gets \omega_{g} - \frac{\lambda}{N_\beta+\gamma}\sum_{i=1}^N \mathbb{I}_{\beta = \rho_i} \cdot \nabla_{\omega_{g}} \mathcal{L}\big(y_i, \hat{y}_i\big)
\end{equation*}
where $N_{\beta} = \sum_{i=1}^N \mathbb{I}_{\beta = \rho_i}$.
} 
\end{algorithm}

\subsection{End-to-End Training in Orchestrated Systems}

In an orchestrated setting, a payments network is responsible for transactions across a pool of banks, and for monitoring potential fraud. It is feasible to facilitate end-to-end training of a privacy-preserving federated pipeline, using model partitioning with back-propagation. To that end, model layers are distributed across data owners:
\begin{itemize}
    \item The \textbf{orchestrating network} owns the scoring function; and supplies banks with information to train their models.
    \item \textbf{Banks} own independent representation learning modules, with custom architectures and data inputs.
\end{itemize}
Information shared relates to locally-differentially private profiles and model gradients; as shown in a naive mini-batch step with \textit{stochastic gradient descent} in Algorithm \ref{alg:orch}. The training process relies on Jacobian-vector products for efficient computations in \eqref{eq:chain}, as well as micro-batches and pipeline parallelism \cite{huang2019gpipe, 10.1145/3341301.3359646} to reduce memory demand and avoid idle processes. We note that the chain rule in \ref{eq:chain} ignores any noise added within additive mechanisms for data publication in \ref{eq:mechanism}, due to simple reparametrisation rules \cite{kingma2015variational}. Here, the algorithm scales with a complexity of $O(B N^2 \omega)$ in a mini-batch of size $N$, where $w$ denotes the maximum number of weights across scoring function and data-release mechanisms.
\begin{algorithm}[h!]
\caption{SGD step. End-to-End Orchestrated.}\label{alg:orch}
\KwIn{Mini-batch $\{(\boldsymbol{x}_{i, t_i}, y_i)\}_{i=1,\dots, N}$. \newline Loss function $\mathcal{L}$ and learning rate $\lambda>0$. \newline Slack term $\gamma\rightarrow0^+$.
}
\KwOut{Updated $s$ and $\psi_\beta$, $\beta=1,\dots, B$.}
\For{$i=1, \dots, N$}{
    $\eta_i \gets getOrdering(\boldsymbol{x}_{i, t_i})$ \\
    $\rho_i \gets getBeneficiary(\boldsymbol{x}_{i, t_i})$ \\
    $\hookrightarrow$ \hfill \textbf{Ordering Bank } $\eta_i$ \\
    \hfill Publish $\boldsymbol{z}^{\text{o}}_{i, t_i} = \mathcal{M}_{\eta_i}(\boldsymbol{X}^{\text{O}}_{\leq t_i})$ \\
    $\hookrightarrow$ \hfill \textbf{Beneficiary Bank } $\rho_i$ \\
    \hfill Publish $\boldsymbol{z}^{\text{b}}_{i, t_i} = \mathcal{M}_{\rho_i}(\boldsymbol{X}^{\text{B}}_{\leq t_i})$ \\
    Predict $\hat{y}_i = s(\boldsymbol{z}^{\text{o}}_{i, t_i}, \boldsymbol{z}^{\text{b}}_{i, t_i}, \boldsymbol{x}_{i, t_i})$
}
Update weights $\omega_s$ of $s$:
\begin{equation*}
    \omega_s \gets \omega_s - \frac{\lambda}{N}\sum_{i=1}^N \nabla_{\omega_s}\mathcal{L}\big(y_i, \hat{y}_i\big)
\end{equation*}
\For(update weights $\omega_{\psi}$ of $\psi_{\beta}$:){$\beta=1,\dots, B$}{
    \For(\textbf{Orchestrator} publish:){$i=1, \dots, N$}{
        \begin{equation*}
            \frac{\partial\mathcal{L}}{\partial\boldsymbol{z}_{i,t_i}} = \big[\mathbb{I}_{\beta=\eta_i} \cdot \nabla_{\boldsymbol{z}^{\text{o}}} + \mathbb{I}_{\beta=\rho_i} \cdot \nabla_{\boldsymbol{z}^{\text{b}}} \big] \mathcal{L}\big(y_i,\hat{y}_i\big)
        \end{equation*}
    }
    update
    \begin{equation}
        \omega_{\psi} \gets \omega_{\psi} - \frac{\lambda}{N_\beta+\gamma} \sum_{i=1}^N \frac{\partial\mathcal{L}}{\partial\boldsymbol{z}_{i,t_i}} \frac{\Psi_{\beta,i}}{\partial\omega_\psi} \label{eq:chain}
    \end{equation}

    where
    \begin{equation*}
        \Psi_{\beta,i} = 
        \mathbb{I}_{\beta=\eta_i} \cdot \psi_\beta(\boldsymbol{X}^O_{\leq t_i}) + \mathbb{I}_{\beta=\rho_i} \cdot \psi_\beta(\boldsymbol{X}^B_{\leq t_i})
    \end{equation*}    

    and $N_\beta = \sum_{i=1}^N \big(\mathbb{I}_{\beta=\eta_i} + \mathbb{I}_{\beta=\rho_i}\big)$.
}
\end{algorithm}

\subsection{Threat Models and Privacy Attacks} \label{sec:attacks}

We build over inference-time attack taxonomies designed for language models \cite{song2020information}, and work to exploit transaction profiles published by external institutions. The goal is to reverse engineer sensitive inputs, as well as drawing private inferences and establishing train data membership. For the purpose, we mimic popular implementations \cite{shokri2017membership, fredrikson2015model, xu2021privacy, liu2022threats}, and pose an elevated threat model. Here, a malicious actor is allowed exposure to a large auxiliary subset of real data in order to efficiently train the attacks (cf. \cite{song2020information, shokri2017membership}). 

\subsubsection{Embedding Inversion Attack}

We invert the embeddings published by external institutions, and aim to recover raw inputs in transaction sequences for accounts. We leverage a black box scenario, using profiles $\boldsymbol{z}_{t} = \mathcal{M}^{\epsilon}_{\text{Laplace}}(\psi(\boldsymbol{X}_{\leq t}))$ and inputs $\boldsymbol{X}_{\leq t}$. An attacker trains an inversion model $\Upsilon:\mathbb{R}^m\rightarrow\mathcal{X}$ with mean squared error loss
\begin{equation*}
    \frac{1}{N}\sum_{i=1}^N\rVert \Upsilon(\boldsymbol{z}_{i, t_i}) - \boldsymbol{x}_{i, t_i} \rVert^2_2
\end{equation*}
for numerical data, using a deep network and back-propagation across auxiliary tuples $\{(\boldsymbol{X}_{i, \leq t_i}, \boldsymbol{z}_{i, t_i})\}_{i=1,\dots,N}$. Thus, we reconstruct the final transaction in each sequence. In our experiments, we report $R^2$ metrics averaged across features in $\mathcal{X}$, measured on validation data hidden to the attacker.

\subsubsection{Attribute Inference Attack}

We infer sensitive attributes about transacting accounts, beyond any information already encoded in embedded profiles. This may include demographic data within identifiers for regions, banks or cards. We use embedded profiles along with account attributes. For each attribute class $a\in\mathcal{A}$, the attacker trains a multinomial classifier $\Upsilon:\mathbb{R}^m\rightarrow \Delta^{|\mathcal{A}|-1}$ with cross-entropy loss
\begin{equation*}
    -\frac{1}{N}\sum_{i=1}^N \log \Upsilon_{a_i}(\boldsymbol{z}_{i, t_i})
\end{equation*}
using auxiliary tuples $\{(\boldsymbol{z}_{i, t_i}, a_i)\}_{i=1,\dots,N}$. This maps embeddings to membership probabilities in a $(|\mathcal{A}|-1)$-simplex. Intuitively, the attacker correlates published profiles with population level characteristics for accounts. This is possible because similar accounts are embedded in proximity to each other. In our experiments, we report on predictive accuracy.

\subsubsection{Membership Attack}

We determine whether personal accounts were used to train a federated scoring function. Here, we use embedded profiles along with class labels for train membership $d\in\{0, 1\}$. We further include mean $\mu_s$ and standard deviation $\sigma_s$ values in scores produced by a shadow fraud classifier for each profile \cite{shokri2017membership}. High epistemic uncertainties in classification tasks are known to be representative of hold-out data inputs \cite{perez2022attribution}. The attacker trains a regression function $\Upsilon:\mathbb{R}^{m+2}\rightarrow [0, 1]$ for binary classification, using a cross-entropy loss
\begin{equation*}
    -\frac{1}{N}\sum_{i=1}^N \Big( d_i \cdot \log \upsilon_i + (1-d_i)  \cdot\log (1-\upsilon_i) \Big)
\end{equation*}
with predictions $\upsilon_i=\Upsilon(\boldsymbol{z}_{i, t_i}, \mu_{i,s}, \sigma_{i,s})$ across auxiliary tuples $\{(\boldsymbol{z}_{i, t_i}, \mu_{i,s}, \sigma_{i,s}, d_i)\}_{i=1,\dots,N}$, which include train ($d=1$) and test ($d=0$) data. In our experiments, we report on hold-out predictive accuracy.

\section{Experimental Results} \label{sec:Experiments}

We present privacy-preserving experimentation results on federated classification tasks for anomaly detection and fraud prevention, using real and synthetic payments data in networked systems. This includes synthetic financial transaction data provided by the SWIFT messaging network \cite{SwiftSite} within the Privacy-Enhancing Technologies (PETs) Prize Challenges competition \cite{PETsChallenge}, as well as real payments data from a Western European economy provided by Featurespace \cite{Featurespace} for research purposes. We evaluate key performance metrics and success rates for privacy attacks under the threat model described in Subsection \ref{sec:attacks}. A data access statement is offered in Appendix \ref{sec:dataAccess}; model implementation details and hyper-parameter choices are summarised in Appendix \ref{sec:implementations}.

\subsection{U.K.-U.S. PETs Prize Challenges} \label{sec:expA}

\begin{table*}[t!]
\centering
\caption{Anomaly detection rates for SWIFT transaction data, along with adversarial attack performance on embeddings.} \label{tab:swiftresults}
\setlength{\tabcolsep}{7pt}
\renewcommand{\arraystretch}{1.15}

\resizebox{0.98\textwidth}{!}{
\begin{tabular}{|l||c|cccc||c|c|c|} \hline
                           & \textbf{AUC} &              & \multicolumn{3}{c||}{\textbf{TPR@FPR}} & \textbf{Inversion} & \textbf{Membership} & \textbf{Inference} \\
\textbf{Budget-$\epsilon$} & PR           & \textbf{FPR} & 1\%         & 0.5\%      & 0.1\%        & Mean-$R^2$         & F-Score             & Accuracy           \\ \hline
$\rightarrow 0^+$          & 81.53 \textit{(0.02)} &              & 96.57 \textit{(0.07)} & 96.32 \textit{(0.07)} & 95.80 \textit{(0.08)} & 0.00 \textit{(0.000)}       & 0.66 \textit{(0.001)}        & 0.25 \textit{(0.003)}       \\
0.5                        & 83.88 \textit{(0.02)} &              & 96.70 \textit{(0.07)} & 96.50 \textit{(0.07)} & 96.05 \textit{(0.07)} & 0.00 \textit{(0.000)}       & 0.66 \textit{(0.001)}        & 0.25 \textit{(0.003)}       \\
1.0                        & 86.01 \textit{(0.03)} &              & 96.81 \textit{(0.07)} & 96.62 \textit{(0.07)} & 96.23 \textit{(0.06)} & 0.00 \textit{(0.000)}       & 0.66 \textit{(0.001)}        & 0.26 \textit{(0.002)}       \\
2.0                        & 87.68 \textit{(0.04)} &              & 96.97 \textit{(0.08)} & 96.77 \textit{(0.07)} & 96.39 \textit{(0.06)} & 0.00 \textit{(0.000)}       & 0.67 \textit{(0.003)}        & 0.26 \textit{(0.002)}       \\
3.0                        & 88.88 \textit{(0.04)} &              & 97.26 \textit{(0.08)} & 97.04 \textit{(0.07)} & 96.68 \textit{(0.06)} & 0.00 \textit{(0.000)}       & 0.69 \textit{(0.007)}        & 0.27 \textit{(0.003)}       \\
5.0                        & 89.84 \textit{(0.03)} &              & 97.51 \textit{(0.07)} & 97.31 \textit{(0.07)} & 96.99 \textit{(0.06)} & 0.01 \textit{(0.002)}       & 0.72 \textit{(0.015)}        & 0.27 \textit{(0.004)}       \\
10.0                       & 90.39 \textit{(0.03)} &              & 97.66 \textit{(0.07)} & 97.46 \textit{(0.06)} & 97.15 \textit{(0.06)} & 0.09 \textit{(0.011)}       & 0.75 \textit{(0.025)}        & 0.29 \textit{(0.009)}       \\ \hline
\textbf{Centralised}       & 90.65 \textit{(0.03)} &              & 97.76 \textit{(0.07)} & 97.55 \textit{(0.06)} & 97.24 \textit{(0.06)} & 0.25 \textit{(0.029)}       & 0.77 \textit{(0.033)}        & 0.33 \textit{(0.019)}       \\ \hline 
\end{tabular}}
\end{table*}

We present an awarded submission to the financial track in the Privacy-Enhancing Technologies Prize Challenges \cite{PETsChallenge}, set by governmental bodies in the U.K. and U.S. in order to mature collaborative learning and privacy-preserving solutions. The goal is to identify anomalous transactions in synthetic data donated by the SWIFT messaging provider, without compromising private account data for transacting accounts. Here, transactions represent payments from ordering to beneficiary banks across multiple financial jurisdictions, and contain data elements as defined in the ISO20022-pacs008 messaging format \cite{iso}. This includes a transaction \textit{reference}, \textit{time stamp} and instructed or settlement \textit{amounts} with \textit{currency} types, along with identifiers for ordering and beneficiary \textit{accounts} as well as banking \textit{institutions}. Finally, there exist \textit{anomaly} labels representing a dependent variable for classification. 
\begin{figure}[h]
    \centering
    \includegraphics[width=0.48\textwidth]{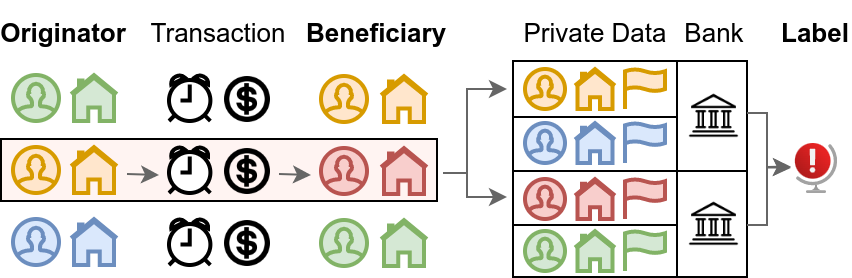}
    \caption{Diagram of the PETs Prize Challenges setting. To explore anomaly detection, SWIFT transaction data may be augmented by requesting personal account information to banking institutions.}
    \label{fig:diagramSWIFT}
\end{figure}

In order to facilitate a collaborative learning scenario, we are further provided with account-level personal data hosted by banks, which includes \textit{names}, \textit{addresses} and \textit{regional codes}, as well as \textit{flags} that outline issues or special account features, such as suspensions. Personal data can be cross-referenced with transactions through the unique identifiers for accounts, as shown in Figure \ref{fig:diagramSWIFT}, and we may compare names and address details with those made public to SWIFT during transactions. We analyse a total of $3997544$ transactions involving over $100$ jurisdictions across train and validation data; using a 75\%-25\% split. We observe $860921$ unique accounts transacting in $59$ currency types, and an anomaly rate of $0.12\%$.

\noindent\textbf{Setting}. We leverage a privacy-preserving orchestrated model with an end-to-end training regime as introduced in Algorithm \ref{alg:orch}. Here, SWIFT acts as an orchestrating network that issues personal data requests to banks, which hold a range of numerically engineered features for accounts under their management. The features define user profiles from categorical flag information, which is highly predictive of anomalous behaviour, and further encode name and address details using writing scripts, character encodings, overall lengths and proportions of vowels, consonants and numbers. However, all features must remain private. Upon request, a bank will process the account information through an embedding model, and publish its output using an additive mechanism endowing local differential privacy.

\noindent\textbf{Predictive performance and privacy attacks}. In Table \ref{tab:swiftresults} we observe average performance metrics for anomaly detection on validation data, along with standard errors computed over randomness in $20$ train and evaluation experiments for each level of privacy budget $\epsilon$. Budgets are used for data publication in additive mechanisms \eqref{eq:mechanism} and assumed consistent across all banks. Metrics include the area under the \textit{Precision-Recall} (PR) curve and \textit{True Positive Rates} (TPR) at reference \textit{False Positive Rate} (FPR) thresholds. Larger budgets are associated with noticeable uptakes in predictive performance, and we notice convergence towards performance in a fully centralised system. Also, we show success rates for inference-time attacks under the elevated threat model introduced in Subsection \ref{sec:attacks}. Here, inference attacks aim to correlate the private embedded data with the categorical ordering bank accounts. Overall, attack performance marginally improves with increasing values of $\epsilon$, compared to (extremely secure) reference values using negligible budgets approaching $0$, with mild improvements observed only for membership attacks. It is thus possible to encode flag statuses for personal accounts while hindering reverse engineering efforts, as well as facilitating entity resolution for names and addresses across account and transaction data, using numerically encoded feature representations.

\subsection{Real-World Acquiring Institution} \label{sec:expB}

\begin{figure}[h!]
    \centering
    \includegraphics[width=0.45\textwidth]{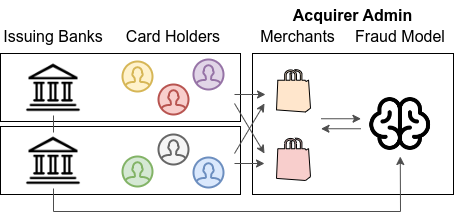}
    \caption{Diagram of collaborative setting. Card holders transact with merchants managed by the acquirer, which accepts responsibility to assess fraud risk. Issuing institutions agree to share privacy-preserving insights for card holders.}
    \label{fig:diagramColoso}
\end{figure}

 Next, we explore real payments data in a transaction network that services a Western European economy. Payments are initiated by cards issued by over 8 thousand retail banks, and credit over 6 million unique merchants. All merchants are serviced by 136 different FIs referred to as \textit{acquirers}, whose responsibility is to accept and process payments. A payment contains information including the \textit{time}, \textit{monetary amount} and \textit{currency}, as well as identifiers for \textit{issuing}, \textit{acquiring}, \textit{card} and \textit{merchant} accounts; also, it is marked as \textit{fraudulent} if there exists a chargeback claim against it. Our experiment takes inspiration from fraud detection systems maintained by acquirers. These receive waivers from \textit{strong authentication} procedures enforced by issuing institutions \cite{fcaReg}, provided fraud rates are low. We explore the scenario pictured in Figure \ref{fig:diagramColoso}, where issuing banks collaborate in order to improve detection rates in a system hosted by an acquirer. To this end, they service privacy-preserving transaction profiles for cards initiating transactions with its merchants. 

\begin{figure*}[t]
    \centering
    \includegraphics[width=0.98\textwidth]{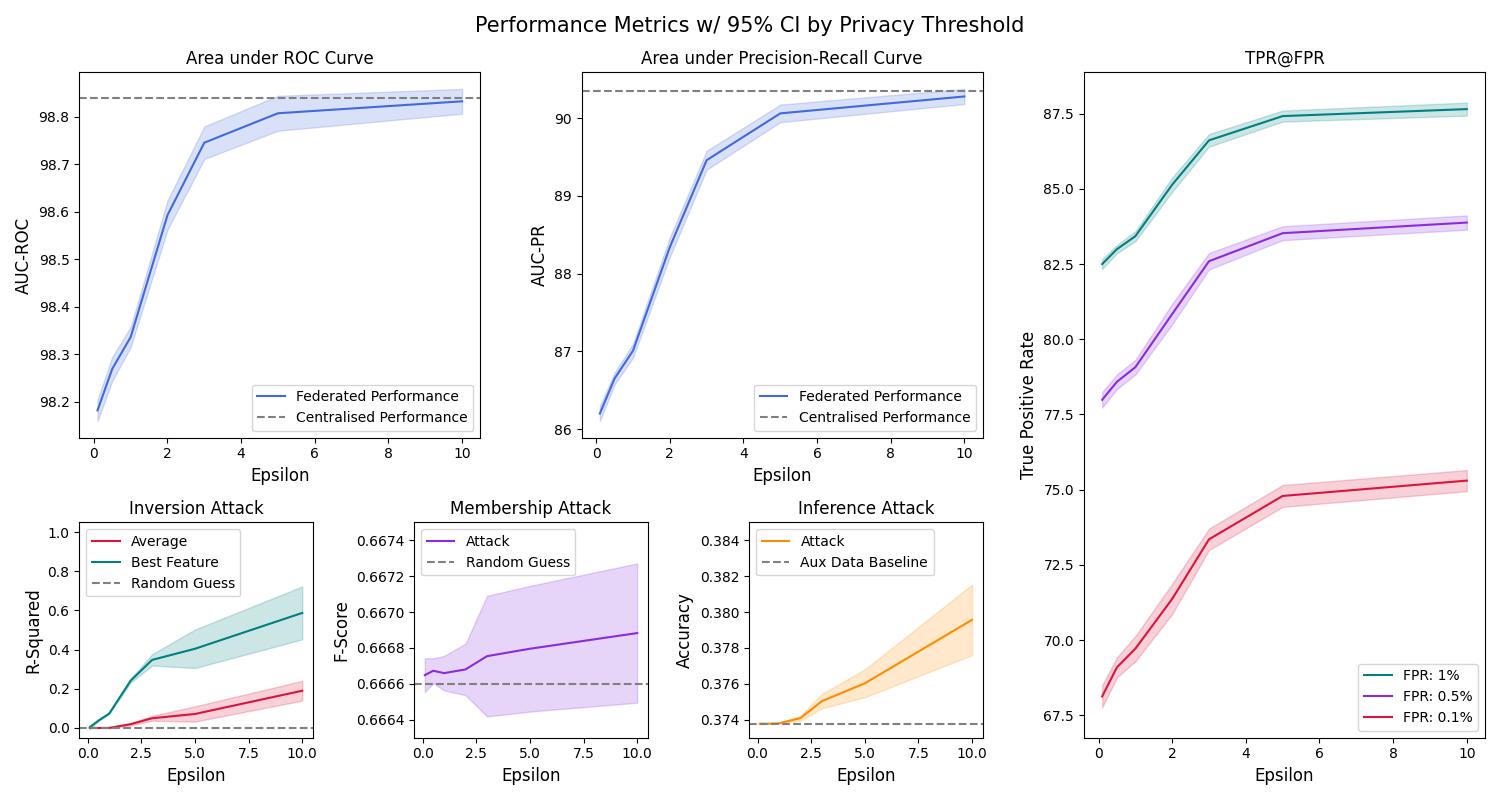}
    \caption{Fraud detection performance metrics and privacy attack success rates for a large acquiring institution, evaluated on a hold-out period.}
    \label{fig:realResults}
\end{figure*}

\noindent\textbf{Setting}. We leverage a privacy-preserving peer-to-peer model with pre-trained embeddings, as introduced in Algorithm \ref{alg:peer}. Here, the purchasing habits for card holders are featurised \textit{per issuing institution} (i.e. data owners), using a combination of unsupervised recurrent architectures and manual processes; and we refer the reader to \cite{micci2001preprocessing, bahnsen2016feature} for example feature engineering strategies. We focus our analysis on the largest acquiring institution in the network, which services over 300,000 merchants; table \ref{tab:realExp} offers a summary of its transaction volumes and fraud prevalence in one year. In order to inform its fraud model, this acquirer issues requests for card profiles in transactions involved with merchants under its management.
\begin{table}[h!]
\centering
\caption{Transaction volumes and banks represented in the data.} \label{tab:realExp}
\setlength{\tabcolsep}{4pt}
\renewcommand{\arraystretch}{1.15}
\resizebox{0.48\textwidth}{!}{
\begin{tabular}{|l|cccc|}
\hline
 & \textbf{Acquirers} & \textbf{Issuers} & \textbf{Transactions}  & \textbf{Fraud} \% \\ \hline
Data Totals   & 136 & 8494 & 5.72×$10^9$ & 0.0132 \\
Target Acquirer &   & 8014 & 1.25×$10^9$ & 0.0076 \\ \hline
\end{tabular}}
\end{table}

\noindent\textbf{Predictive performance and privacy attacks}. In Figure \ref{fig:realResults} we visualise key metrics along with $95\%$ confidence intervals, evaluated on 25\% hold-out data. Results are produced over $20$ repeated train and evaluation experiments at each privacy budget $\epsilon$. Average metrics are further summarised in Table \ref{tab:realResults}, including the area under the \textit{Receiver Operating Characteristic} (ROC) and PR curves, as well as TPR rates at reference FPR thresholds. In all cases, the horizontal axis represents a varying budget $\epsilon$ in additive mechanisms \eqref{eq:mechanism}. Large budgets are associated with a noticeable uptake in predictive performance, and trends converge towards benchmark values (dashed horizontal lines) representing a fully centralised system. Additionally, we explore success rates for inference-time \textit{inversion}, \textit{membership} and \textit{inference} attacks against published embeddings, under an elevated threat model as introduced in Subsection \ref{sec:attacks}. Specifically, inference attacks aim to predict the issuing bank and card \textit{identifiers} for transactions. We compare attack performance versus benchmarks determined by random guessing or using frequencies observed in auxiliary data. We note that both membership and inference attacks are unsuccessful even in high-budget scenarios, with performance marginally improved over benchmarks. On the other hand, inversion attacks show limited success in low privacy and fully centralised settings. Overall, utility-to-privacy profiles resemble similar analyses on different application domains \cite{pan2020privacy} with limited leakage of precise transacting activity information for card holders.
\begin{table}[b!]
\centering
\caption{Mean predictive and adversarial attack performance.} \label{tab:realResults}
\setlength{\tabcolsep}{6pt}
\renewcommand{\arraystretch}{1.15}

\resizebox{0.48\textwidth}{!}{
\begin{tabular}{|l|cc|cccc|}
\hline
\textbf{}            & \multicolumn{2}{c|}{\textbf{AUC}} & & \multicolumn{3}{c|}{\textbf{TPR@FPR}} \\
\textbf{Budget}-$\boldsymbol{\epsilon}$ & ROC  & PR & \textbf{FPR} & 1\%  & 0.5\% & 0.1\% \\ \hline
$\rightarrow 0^+$    & 98.18 & 86.19 & & 82.50 & 77.98 & 68.12 \\
0.5                  & 98.26 & 86.65 & & 82.99 & 78.58 & 69.09 \\
1.0                  & 98.33 & 87.00 & & 83.42 & 79.07 & 69.71 \\
2.0                  & 98.59 & 88.34 & & 85.13 & 80.83 & 71.36 \\
3.0                  & 98.74 & 89.46 & & 86.61 & 82.59 & 73.34 \\
5.0                  & 98.80 & 90.06 & & 87.41 & 83.52 & 74.79 \\
10.0                 & 98.83 & 90.28 & & 87.65 & 83.87 & 75.30 \\ \hline
\textbf{Centralised} & 98.84 & 90.35 & & 87.69 & 84.00 & 75.48 \\ \hline
\end{tabular}}
\\[0.25cm]

\setlength{\tabcolsep}{5pt}
\resizebox{0.48\textwidth}{!}{
\begin{tabular}{|l|cc|c|c|}
\hline
\textbf{}                                & \multicolumn{2}{c|}{\textbf{Inversion}} & \textbf{Membership} & \textbf{Inference} \\
\textbf{Budget}-$\boldsymbol{\epsilon}$  & Mean-$R^2$ & Best-$R^2$ & F-Score & Accuracy \\ \hline
$\rightarrow 0^+$                        & 0.00 & 0.00 & 0.67 & 0.37 \\
0.5                                      & 0.00 & 0.03 & 0.67 & 0.37 \\
1.0                                      & 0.00 & 0.07 & 0.67 & 0.37 \\
2.0                                      & 0.01 & 0.24 & 0.67 & 0.37 \\
3.0                                      & 0.04 & 0.34 & 0.67 & 0.38 \\
5.0                                      & 0.07 & 0.40 & 0.67 & 0.38 \\
10.0                                     & 0.19 & 0.58 & 0.67 & 0.38 \\ \hline
\textbf{Centralised}                     & 0.40 & 0.84 & 0.67 & 0.39 \\ \hline
\textbf{Benchmark}                       & 0.00 & 0.00 & 0.67 & 0.37 \\ \hline
\end{tabular}}
\end{table}

\section{Conclusion} \label{sec:Conclusion}
In this paper, we have introduced a novel framework for privacy-preserving collaborative learning, which relies on embedding models \cite{gui2016large, reimers2019sentence, song2020information} and differential privacy techniques \cite{dwork2014algorithmic}. We have developed a secure data publication mechanism which can inform externally hosted fraud and anomaly detection systems, and presented experimental results on synthetic and real world payments data sets. We have shown utility-privacy profiles analogous to results in privacy-preserving applications across different domains \cite{mahendran2015understanding, coavoux2018privacy, ulyanov2018deep, song2020information}, and considered popular threat models with inversion, membership and attribute inference attacks. Thus, we have offered evidence of the potential to facilitate collaborative solutions for important problems in the financial sector, without requirements to over-share the sensitive data of individuals, banking institutions or merchants.






\bibliographystyle{IEEEtran}
\bibliography{bibliography}

\begin{thebibliography}{10}
\providecommand{\url}[1]{#1}
\csname url@samestyle\endcsname
\providecommand{\newblock}{\relax}
\providecommand{\bibinfo}[2]{#2}
\providecommand{\BIBentrySTDinterwordspacing}{\spaceskip=0pt\relax}
\providecommand{\BIBentryALTinterwordstretchfactor}{4}
\providecommand{\BIBentryALTinterwordspacing}{\spaceskip=\fontdimen2\font plus
\BIBentryALTinterwordstretchfactor\fontdimen3\font minus
  \fontdimen4\font\relax}
\providecommand{\BIBforeignlanguage}[2]{{%
\expandafter\ifx\csname l@#1\endcsname\relax
\typeout{** WARNING: IEEEtran.bst: No hyphenation pattern has been}%
\typeout{** loaded for the language `#1'. Using the pattern for}%
\typeout{** the default language instead.}%
\else
\language=\csname l@#1\endcsname
\fi
#2}}
\providecommand{\BIBdecl}{\relax}
\BIBdecl

\bibitem{unCrime}
\BIBentryALTinterwordspacing
{United Nations}, ``{Money Laundering},'' accessed on Jun 1st, 2023. [Online].
  Available: \url{www.unodc.org/unodc/en/money-laundering/overview.html}
\BIBentrySTDinterwordspacing

\bibitem{FATF}
\BIBentryALTinterwordspacing
{Financial Action Task Force}, ``{Stocktake on Data Pooling, Collaborative
  Analytics and Data Protection},'' accessed on Jun 1st, 2023. [Online].
  Available:
  \url{www.fatf-gafi.org/publications/digitaltransformation/documents/data-pooling-collaborative-analytics-data-protection.html}
\BIBentrySTDinterwordspacing

\bibitem{PETsChallenge}
\BIBentryALTinterwordspacing
``{U.K.-U.S.} {P}rivacy {E}nhancing {T}echnologies {(PETs)} prize challenges,''
  accessed on Jun 1st, 2023. [Online]. Available: \url{petsprizechallenges.com}
\BIBentrySTDinterwordspacing

\bibitem{roy2018deep}
A.~Roy, J.~Sun, R.~Mahoney, L.~Alonzi, S.~Adams, and P.~Beling, ``Deep learning
  detecting fraud in credit card transactions,'' in \emph{Systems and
  Information Engineering Design Symposium (SIEDS)}.\hskip 1em plus 0.5em minus
  0.4em\relax IEEE, 2018.

\bibitem{branco2020interleaved}
B.~Branco, P.~Abreu, A.~S. Gomes, M.~S. Almeida, J.~T. Ascens{\~a}o, and
  P.~Bizarro, ``Interleaved sequence {RNN}s for fraud detection,'' in
  \emph{26th ACM SIGKDD international conference on knowledge discovery \& data
  mining}, 2020.

\bibitem{wong2023training}
K.~Wong, D.~Sutton, I.~Perez, and A.~Barns-Graham, ``Training a machine
  learning system for transaction data processing,'' 2023, {US} Patent App.
  17/420,159.

\bibitem{hochreiter1997long}
S.~Hochreiter and J.~Schmidhuber, ``Long short-term memory,'' \emph{Neural
  computation}, vol.~9, 1997.

\bibitem{yu2019review}
Y.~Yu, X.~Si, C.~Hu, and J.~Zhang, ``A review of recurrent neural networks:
  Lstm cells and network architectures,'' \emph{Neural computation}, vol.~31,
  2019.

\bibitem{9912385}
Y.~Xie, G.~Liu, C.~Yan, C.~Jiang, M.~Zhou, and M.~Li, ``Learning transactional
  behavioral representations for credit card fraud detection,''
  \emph{Transactions on Neural Networks and Learning Systems}, 2022.

\bibitem{vaswani2017attention}
A.~Vaswani, N.~Shazeer, N.~Parmar, J.~Uszkoreit, L.~Jones, A.~N. Gomez,
  {\L}.~Kaiser, and I.~Polosukhin, ``Attention is all you need,''
  \emph{Advances in neural information processing systems, NIPS}, 2017.

\bibitem{lecun2015deep}
Y.~LeCun, Y.~Bengio, and G.~Hinton, ``Deep learning,'' \emph{Nature}, vol. 521,
  2015.

\bibitem{shokri2015privacy}
R.~Shokri and V.~Shmatikov, ``Privacy-preserving deep learning,'' in \emph{22nd
  ACM SIGSAC conference on computer and communications security}, 2015.

\bibitem{nasr2018comprehensive}
M.~Nasr, R.~Shokri, and A.~Houmansadr, ``Comprehensive privacy analysis of deep
  learning,'' in \emph{Symposium on Security and Privacy (SP)}.\hskip 1em plus
  0.5em minus 0.4em\relax IEEE, 2018.

\bibitem{carlini2019secret}
N.~Carlini, C.~Liu, {\'U}.~Erlingsson, J.~Kos, and D.~Song, ``The secret
  sharer: Evaluating and testing unintended memorization in neural networks.''
  in \emph{USENIX Security Symposium}, vol. 267, 2019.

\bibitem{zhu2019deep}
L.~Zhu, Z.~Liu, and S.~Han, ``Deep leakage from gradients,'' \emph{Advances in
  neural information processing systems, NIPS}, 2019.

\bibitem{yin2021comprehensive}
X.~Yin, Y.~Zhu, and J.~Hu, ``A comprehensive survey of privacy-preserving
  federated learning: A taxonomy, review, and future directions,'' \emph{ACM
  Computing Surveys}, vol.~54, 2021.

\bibitem{mothukuri2021survey}
V.~Mothukuri, R.~M. Parizi, S.~Pouriyeh, Y.~Huang, A.~Dehghantanha, and
  G.~Srivastava, ``A survey on security and privacy of federated learning,''
  \emph{Future Generation Computer Systems}, vol. 115, 2021.

\bibitem{lyu2020threats}
L.~Lyu, H.~Yu, J.~Zhao, and Q.~Yang, ``Threats to federated learning,''
  \emph{Federated Learning: Privacy and Incentive}, 2020.

\bibitem{coavoux2018privacy}
M.~Coavoux, S.~Narayan, and S.~B. Cohen, ``Privacy-preserving neural
  representations of text,'' \emph{arXiv preprint arXiv:1808.09408}, 2018.

\bibitem{mahendran2015understanding}
A.~Mahendran and A.~Vedaldi, ``Understanding deep image representations by
  inverting them,'' in \emph{Conference on computer vision and pattern
  recognition}.\hskip 1em plus 0.5em minus 0.4em\relax IEEE, 2015.

\bibitem{ulyanov2018deep}
D.~Ulyanov, A.~Vedaldi, and V.~Lempitsky, ``Deep image prior,'' in
  \emph{Conference on computer vision and pattern recognition}.\hskip 1em plus
  0.5em minus 0.4em\relax IEEE, 2018.

\bibitem{song2020information}
C.~Song and A.~Raghunathan, ``Information leakage in embedding models,'' in
  \emph{2020 ACM SIGSAC conference on computer and communications security},
  2020.

\bibitem{dwork2008differential}
C.~Dwork, ``Differential privacy: A survey of results,'' in \emph{Theory and
  Applications of Models of Computation: 5th International Conference, TAMC
  2008, Xi’an, China, April 25-29, 2008. Proceedings 5}, 2008.

\bibitem{abadi2016deep}
M.~Abadi, A.~Chu, I.~Goodfellow, H.~B. McMahan, I.~Mironov, K.~Talwar, and
  L.~Zhang, ``Deep learning with differential privacy,'' in \emph{ACM SIGSAC
  conference on computer and communications security}, 2016.

\bibitem{aono2017privacy}
Y.~Aono, T.~Hayashi, L.~Wang, S.~Moriai \emph{et~al.}, ``Privacy-preserving
  deep learning via additively homomorphic encryption,'' \emph{Transactions on
  Information Forensics and Security}, vol.~13, 2017.

\bibitem{knott2021crypten}
B.~Knott, S.~Venkataraman, A.~Hannun, S.~Sengupta, M.~Ibrahim, and L.~van~der
  Maaten, ``Crypten: Secure multi-party computation meets machine learning,''
  \emph{Advances in Neural Information Processing Systems, NIPS}, 2021.

\bibitem{SwiftSite}
\BIBentryALTinterwordspacing
``{Society for Worldwide Interbank Financial Telecommunication},'' accessed on
  Jun 1st, 2023. [Online]. Available: \url{www.swift.com}
\BIBentrySTDinterwordspacing

\bibitem{liu2020asymmetrical}
Y.~Liu, X.~Zhang, and L.~Wang, ``Asymmetrical vertical federated learning,''
  \emph{arXiv preprint arXiv:2004.07427}, 2020.

\bibitem{romanini2021pyvertical}
D.~Romanini, A.~J. Hall, P.~Papadopoulos, T.~Titcombe, A.~Ismail, T.~Cebere,
  R.~Sandmann, R.~Roehm, and M.~A. Hoeh, ``Pyvertical: A vertical federated
  learning framework for multi-headed splitnn,'' \emph{arXiv preprint
  arXiv:2104.00489}, 2021.

\bibitem{melis2019exploiting}
L.~Melis, C.~Song, E.~De~Cristofaro, and V.~Shmatikov, ``Exploiting unintended
  feature leakage in collaborative learning,'' in \emph{Symposium on security
  and privacy (SP)}.\hskip 1em plus 0.5em minus 0.4em\relax IEEE, 2019.

\bibitem{Featurespace}
\BIBentryALTinterwordspacing
``Featurespace,'' accessed on Jun 1st, 2023. [Online]. Available:
  \url{www.featurespace.co.uk}
\BIBentrySTDinterwordspacing

\bibitem{bolton2002statistical}
R.~J. Bolton and D.~J. Hand, ``Statistical fraud detection: A review,''
  \emph{Statistical science}, vol.~17, 2002.

\bibitem{iso}
\BIBentryALTinterwordspacing
{Universal Financial Industry Message Scheme}, ``{ISO 20022 Message
  Definitions},'' accessed on Jun 1st, 2023. [Online]. Available:
  \url{www.iso20022.org/iso-20022-message-definitions}
\BIBentrySTDinterwordspacing

\bibitem{iso2}
\BIBentryALTinterwordspacing
{International Organization for Standardization}, ``{ISO 8583 Message
  Definitions},'' accessed on Jun 1st, 2023. [Online]. Available:
  \url{www.iso.org/obp/ui/#iso:std:iso:8583}
\BIBentrySTDinterwordspacing

\bibitem{hancock2020survey}
J.~T. Hancock and T.~M. Khoshgoftaar, ``Survey on categorical data for neural
  networks,'' \emph{Journal of Big Data}, 2020.

\bibitem{weiss2016survey}
K.~Weiss, T.~M. Khoshgoftaar, and D.~Wang, ``A survey of transfer learning,''
  \emph{Journal of Big data}, 2016.

\bibitem{li2022approximation}
Z.~Li, J.~Han, E.~Weinan, and Q.~Li, ``Approximation and optimization theory
  for linear continuous-time recurrent neural networks.'' \emph{Journal of
  Machine Learning Research}, vol.~23, 2022.

\bibitem{bonawitz2019towards}
K.~Bonawitz, H.~Eichner, W.~Grieskamp, D.~Huba, A.~Ingerman, V.~Ivanov,
  C.~Kiddon, J.~Kone{\v{c}}n{\`y}, S.~Mazzocchi, B.~McMahan \emph{et~al.},
  ``Towards federated learning at scale: System design,'' \emph{Proceedings of
  machine learning and systems}, vol.~1, 2019.

\bibitem{li2020federated}
T.~Li, A.~K. Sahu, A.~Talwalkar, and V.~Smith, ``Federated learning:
  Challenges, methods, and future directions,'' \emph{IEEE signal processing
  magazine}, vol.~37, 2020.

\bibitem{zhang2021survey}
C.~Zhang, Y.~Xie, H.~Bai, B.~Yu, W.~Li, and Y.~Gao, ``A survey on federated
  learning,'' \emph{Knowledge-Based Systems}, vol. 216, 2021.

\bibitem{naehrig2011can}
M.~Naehrig, K.~Lauter, and V.~Vaikuntanathan, ``Can homomorphic encryption be
  practical?'' in \emph{ACM workshop on Cloud computing security workshop},
  2011.

\bibitem{dwork2014algorithmic}
C.~Dwork, A.~Roth \emph{et~al.}, ``The algorithmic foundations of differential
  privacy,'' \emph{Foundations and Trends in Theoretical Computer Science},
  vol.~9, 2014.

\bibitem{heurix2015taxonomy}
J.~Heurix, P.~Zimmermann, T.~Neubauer, and S.~Fenz, ``A taxonomy for privacy
  enhancing technologies,'' \emph{Computers \& Security}, vol.~53, 2015.

\bibitem{dankar2013practicing}
F.~K. Dankar and K.~El~Emam, ``Practicing differential privacy in health care:
  A review.'' \emph{Transactions on Data Privacy}, vol.~6, 2013.

\bibitem{wang2016differential}
L.~Wang, D.~Zhang, D.~Yang, B.~Y. Lim, and X.~Ma, ``Differential location
  privacy for sparse mobile crowdsensing,'' in \emph{International Conference
  on Data Mining (ICDM)}.\hskip 1em plus 0.5em minus 0.4em\relax IEEE, 2016.

\bibitem{xu2021privacy}
R.~Xu, N.~Baracaldo, and J.~Joshi, ``Privacy-preserving machine learning:
  Methods, challenges and directions,'' \emph{arXiv preprint arXiv:2108.04417},
  2021.

\bibitem{darbha2020privacy}
S.~Darbha and R.~Arora, ``Privacy in cbdc technology,'' Bank of Canada, Tech.
  Rep., 2020.

\bibitem{almuhammadi2004better}
S.~Almuhammadi, N.~T. Sui, and D.~McLeod, ``Better privacy and security in
  e-commerce: using elliptic curve-based zero knowledge proofs,'' in
  \emph{International Conference on e-Commerce Technology}.\hskip 1em plus
  0.5em minus 0.4em\relax IEEE, 2004.

\bibitem{kanamori2022privacy}
S.~Kanamori, T.~Abe, T.~Ito, K.~Emura, L.~Wang, S.~Yamamoto, T.~P. Le, K.~Abe,
  S.~Kim, R.~Nojima \emph{et~al.}, ``Privacy-preserving federated learning for
  detecting fraudulent financial transactions in japanese banks,''
  \emph{Journal of Information Processing}, vol.~30, 2022.

\bibitem{canillas2018exploratory}
R.~Canillas, R.~Talbi, S.~Bouchenak, O.~Hasan, L.~Brunie, and L.~Sarrat,
  ``Exploratory study of privacy preserving fraud detection,'' in \emph{19th
  International Middleware Conference Industry}, 2018.

\bibitem{lin2020protecting}
J.~Lin, J.~Niu, X.~Liu, and M.~Guizani, ``Protecting your shopping preference
  with differential privacy,'' \emph{Transactions on Mobile Computing},
  vol.~20, 2020.

\bibitem{gui2016large}
H.~Gui, J.~Liu, F.~Tao, M.~Jiang, B.~Norick, and J.~Han, ``Large-scale
  embedding learning in heterogeneous event data,'' in \emph{International
  Conference on Data Mining (ICDM)}.\hskip 1em plus 0.5em minus 0.4em\relax
  IEEE, 2016.

\bibitem{reimers2019sentence}
N.~Reimers and I.~Gurevych, ``Sentence-bert: Sentence embeddings using siamese
  bert-networks,'' \emph{arXiv preprint arXiv:1908.10084}, 2019.

\bibitem{logeswaran2018an}
L.~Logeswaran and H.~Lee, ``An efficient framework for learning sentence
  representations,'' in \emph{International Conference on Learning
  Representations, ICLR}, 2018.

\bibitem{nguyen2023meet}
A.~Nguyen, N.~Karampatziakis, and W.~Chen, ``Meet in the middle: A new
  pre-training paradigm,'' \emph{arXiv preprint arXiv:2303.07295}, 2023.

\bibitem{babaev2022coles}
D.~Babaev, N.~Ovsov, I.~Kireev, M.~Ivanova, G.~Gusev, I.~Nazarov, and
  A.~Tuzhilin, ``Coles: Contrastive learning for event sequences with
  self-supervision,'' in \emph{International Conference on Management of Data},
  2022.

\bibitem{wainwright2012privacy}
M.~J. Wainwright, M.~Jordan, and J.~C. Duchi, ``Privacy aware learning,''
  \emph{Advances in Neural Information Processing Systems, NIPS}, 2012.

\bibitem{phan2017adaptive}
N.~Phan, X.~Wu, H.~Hu, and D.~Dou, ``Adaptive laplace mechanism: Differential
  privacy preservation in deep learning,'' in \emph{International Conference on
  Data Mining (ICDM)}.\hskip 1em plus 0.5em minus 0.4em\relax IEEE, 2017.

\bibitem{dwork2006calibrating}
C.~Dwork, F.~McSherry, K.~Nissim, and A.~Smith, ``Calibrating noise to
  sensitivity in private data analysis,'' in \emph{Theory of Cryptography:
  Third Theory of Cryptography Conference, New York}, 2006.

\bibitem{huang2019gpipe}
Y.~Huang, Y.~Cheng, A.~Bapna, O.~Firat, D.~Chen, M.~Chen, H.~Lee, J.~Ngiam,
  Q.~V. Le, Y.~Wu \emph{et~al.}, ``Gpipe: Efficient training of giant neural
  networks using pipeline parallelism,'' \emph{Advances in neural information
  processing systems, NIPS}, 2019.

\bibitem{10.1145/3341301.3359646}
D.~Narayanan, A.~Harlap, A.~Phanishayee, V.~Seshadri, N.~R. Devanur, G.~R.
  Ganger, P.~B. Gibbons, and M.~Zaharia, ``Pipedream: Generalized pipeline
  parallelism for dnn training,'' ser. SOSP '19.\hskip 1em plus 0.5em minus
  0.4em\relax Association for Computing Machinery, 2019.

\bibitem{kingma2015variational}
D.~P. Kingma, T.~Salimans, and M.~Welling, ``Variational dropout and the local
  reparameterization trick,'' \emph{Advances in neural information processing
  systems, NIPS}, vol.~28, 2015.

\bibitem{shokri2017membership}
R.~Shokri, M.~Stronati, C.~Song, and V.~Shmatikov, ``Membership inference
  attacks against machine learning models,'' in \emph{Symposium on security and
  privacy}.\hskip 1em plus 0.5em minus 0.4em\relax IEEE, 2017.

\bibitem{fredrikson2015model}
M.~Fredrikson, S.~Jha, and T.~Ristenpart, ``Model inversion attacks that
  exploit confidence information and basic countermeasures,'' in \emph{22nd ACM
  SIGSAC conference on computer and communications security}, 2015.

\bibitem{liu2022threats}
P.~Liu, X.~Xu, and W.~Wang, ``Threats, attacks and defenses to federated
  learning: issues, taxonomy and perspectives,'' \emph{Cybersecurity}, 2022.

\bibitem{perez2022attribution}
I.~Perez, P.~Skalski, A.~Barns-Graham, J.~Wong, and D.~Sutton, ``Attribution of
  predictive uncertainties in classification models,'' in \emph{Uncertainty in
  Artificial Intelligence (UAI)}, 2022.

\bibitem{fcaReg}
\BIBentryALTinterwordspacing
{UK Financial Conduct Authority}, ``{Strong Customer Authentication and Common
  and Secure Methods of Communication},'' accessed on Jun 1st, 2023. [Online].
  Available: \url{www.handbook.fca.org.uk/techstandards/PS/2021/}
\BIBentrySTDinterwordspacing

\bibitem{micci2001preprocessing}
D.~Micci-Barreca, ``A preprocessing scheme for high-cardinality categorical
  attributes in classification and prediction problems,'' \emph{ACM SIGKDD
  Explorations Newsletter}, vol.~3, 2001.

\bibitem{bahnsen2016feature}
A.~C. Bahnsen, D.~Aouada, A.~Stojanovic, and B.~Ottersten, ``Feature
  engineering strategies for credit card fraud detection,'' \emph{Expert
  Systems with Applications}, vol.~51, 2016.

\bibitem{pan2020privacy}
X.~Pan, M.~Zhang, S.~Ji, and M.~Yang, ``Privacy risks of general-purpose
  language models,'' in \emph{Symposium on Security and Privacy (SP)}.\hskip
  1em plus 0.5em minus 0.4em\relax IEEE, 2020.

\end{thebibliography}

\appendix

\subsection{Data and Code Access Statement} \label{sec:dataAccess}

The first experiment uses synthetic third party data provided by the SWIFT network \cite{SwiftSite}, which authors do not have permission to share under a license agreement, and following data protection policies at \url{https://www.swift.com/about-us/legal/compliance/data-protection-policies}. Please direct enquiries about the data and its access arrangements to \url{https://www.swift.com/contact-us}. For enquires on research code for the PETs Challenge submission \cite{PETsChallenge}, as well as further white-paper documentation and evaluation metrics, please contact \href{mailto:petsprizechallenges@cdei.gov.uk}{petsprizechallenges@cdei.gov.uk}.

Data and research code supporting the second experiment is not publicly available due to commercial agreements and the EU General Data Protection Regulation. For further insights, please contact the authors or \href{mailto:info@featurespace.com}{info@featurespace.com}.

\subsection{Model Implementation Details} \label{sec:implementations}

In both settings, we use training batches of size $1024$ and a learning rate of $0.001$, however, we replace vanilla stochastic gradient descent with the \textit{Adam} optimizer. All experiments are performed on an NVIDIA DGX appliance with eight A100 GPUs, each accounting for 40GB of VRAM. Run-times for training exercises are of approximately $2$ hours per end-to-end training job in Experiment \ref{sec:expA}, and $30$ minutes for peer-to-peer training jobs in Experiment  \ref{sec:expB}.

\noindent\textbf{Temporal aggregation modules}. They follow a vanilla LSTM implementation with hidden layers of dimension $32$. Data publication embeds temporal profiles with a $2$-layer \textit{multi-layer perceptron} (MLP), with sizes $64$/$32$, \textit{relu} activation functions and \textit{dropout}. Output vectors are of dimension $8$, and use \textit{tanh} activation, norm clipping and Laplace noise. All categorical input data is pre-processed through tokenising.

\noindent\textbf{Regression function}. Uses a $3$-layer MLP with sizes $128/64/32$, \textit{relu} activation and \textit{dropout}; followed by a single unit output layer with \textit{sigmoid} activation for scoring.

\noindent\textbf{Embedding pre-processors} Algorithm \ref{alg:peer} use MLP models with a single hidden layer of size $32$ and \textit{relu} activation, followed by an output layer of size $8$ and \textit{tanh} activation.

\noindent\textbf{Jacobian-vector products} in \eqref{eq:chain} use micro-batches of size $8$ to limit memory consumption in pipeline parallelism.

\end{document}